\documentclass{aipproc}

\newcommand{\be}{\begin{equation}}
\newcommand{\ee}{\end{equation}}
\newcommand{\bea}{\begin{eqnarray}}
\newcommand{\eea}{\end{eqnarray}} \newcommand{\nn}{\nonumber}

\newcommand{\de}{\partial}

\layoutstyle{6x9}

\SetInternalRegister\hbadness{8000} 

\newcommand\doingARLO[2][]{%
  \ifx\mmref\undefined #1\else #2\fi
}

\begin{document}
\def\esp #1{e^{\displaystyle{#1}}}
\def\slash#1{\setbox0=\hbox{$#1$}#1\hskip-\wd0\dimen0=5pt\advance
       \dimen0 by-\ht0\advance\dimen0 by\dp0\lower0.5\dimen0\hbox
         to\wd0{\hss\sl/\/\hss}}\def\ink {\int~{d^4k\over (2\pi)^4}~}

\title
      [Deconstructed Higgsless Models]
      {Deconstructed Higgsless Models}



\author{Roberto Casalbuoni}{
  address={Dipartimento di Fisica dell' Universita' di Firenze and Sezione
INFN, Via G. Sansone 1, 50019 Sesto Fiorentino (Firenze), Italy.
E-mail: casalbuoni@fi.infn.it},
  email={casalbuoni@fi.infn.it},
}

\begin{abstract}
We consider the possibility of constructing realistic Higgsless
models within the context of deconstructed or moose models. We show
that the constraints coming from the electro-weak esperimental data
are very severe and that it is very difficult to reconcile them with
the requirement of improving the unitarity bound of the Higgsless
Standard Model. On the other hand, with some fine tuning, a solution
is found by delocalizing the standard fermions along the lattice
line, that is allowing  the fermions to couple to the moose gauge
fields.

\end{abstract}

\date{\today}

\maketitle

\section{1. Higher Dimensional Gauge Theories}

In the past few years a renewal of interest in higher dimensional
theories came out of the possibility of sub-millimiter extra
dimensions due to the softening of gravitational theories in a
subspace \cite{Arkani-Hamed:1998rs,Antoniadis:1998ig}. In this way a
strong gravitational interaction in $D$ space-time dimensions
($D>4$) might give rise to a weak gravitational interaction in the
usual 4 dimensions. If the extra dimensions, $d=D-4$, are
compactified, one gets a relation between the Planck scale $M_D$ in
$D$ dimensions and the four-dimensional one, $M_P$ \be M_P^2=R^d
M_D^{2+d},\ee with $R$ the compactification radius. By choosing
$R\gg M_D^{-1}$ one can make $M_D^2\ll M_P^2$. As an example, with
$M_D=1~TeV$ and $d=2$ one gets $R\approx 0.1~mm$.

\begin{figure}
  \includegraphics[height=.1\textheight]{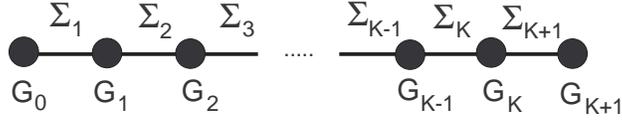}
  \caption{The diagram illustrates a deconstructed theory described by the gauge groups
  $G_i$ and by the chiral fields $\Sigma_i$.\label{fig:2}}
\end{figure}

On the other hand, gauge theories in higher dimensional spaces offer
extra bonus as the possibility of realizing a geometrical Higgs
mechanism. As an example we consider an abelian gauge theory in 4+1
dimensions: \be {\cal L}=-\frac 1{2g_5^2}F_{AB}F^{AB}=-\frac
1{2g_5^2}F_{\mu\nu}F^{\mu\nu}-\frac 1{g_5^2}F_{\mu
5}F^{\mu5}.\label{eq:2}\ee Here $g_5$ is the gauge coupling in 5D
having dimensions of $M^{-1/2}$, $A$, $B$ are the space-time indices
in $D$ dimensions, and $\mu$, $\nu$ the usual 4-dimensional indices.
Furthermore \be F_{AB}=\de_A A_B-\de_B A_A,\ee Performing the gauge
transformation (with the understanding that we omit the zero mode of
the operator $\de_5$)\be A_B\to A_B-(\de_5)^{-1}(\de_B A_5),\ee we
get \be A_5=0 \Rightarrow F_{\mu 5}=-\de_5 A_\mu.\ee If the fifth
dimension is compactified on a circle $S^2$ of length $2\pi R$, the
non zero eigenmodes $A_\mu^n$ of the fields $A_\mu$ acquire a mass
$M_n=n/ R$ since in this case \be A_\mu(x_\mu,x_5)= \sum_n
e^{inx_5/R} A_\mu^n(x_\mu).\ee However the zero mode remains
massless and a GB is present. This zero mode can be eliminated
compactifying the model on an orbifold, that is on the coset
$S^2/Z$, $Z$ being the discrete group of reflections along the fifth
dimension: \be Z:~~x_5\to -x_5.\ee This allows to define fields as
eigenstates of $Z$ \be A_B(x_\mu, x_5)=\pm A_B(x_\mu, -x_5).\ee In
this way various possibilities open up. As an instance, by taking
the odd eigenstates no zero modes are in the spectrum and one gets
only massive gauge bosons. In other words we have obtained massive
gauge bosons in the framework of a gauge theory without Higgs
fields. If the extra dimension is discretized
\cite{Hill:2000mu,Cheng:2001vd} one gets a so-called deconstructed
gauge theory \cite{Arkani-Hamed:2001ca,Arkani-Hamed:2001nc}. In this
construction the connection field along the fifth dimension, $A_5$,
gives rise to a non-linear $\sigma-$field. In fact a gauge field is
nothing but a connection, that is a way of relating the phases of
fields at nearby points. Once the space is discretized the
connection goes naturally into a link variable realizing the
parallel transport between two lattice sites. The link variable $
\Sigma_i= e^{-ia A_5^{i-1}}$ satisfies the condition $
\Sigma\Sigma^\dagger =1$ and it can be identified with a chiral
field. In fact, if we consider a non-abelian gauge theory acting on
the five-dimensional space, through discretization of the fifth
dimension we get a discrete infinity of four-dimensional gauge
theories each of them acting at a particular lattice site. It can be
easily seen that the $\Sigma_i$ fields transform according to \be
\Sigma_i\to U_{i-1} \Sigma_i U_i^\dagger,\ee with $U_{i-1}$ and
$U_{i}$ group transformations belonging to the gauge group $G$
located at the lattice sites $i-1$ and $i$ respectively. Then the
covariant derivatives of the chiral fields can be connected with the
field strengths $F_{\mu 5}$ by \be
D_\mu\Sigma_i=\de_\mu\Sigma_i-iA_\mu^{i-1}+i\Sigma_i A_\mu^i\approx
-ia F_{\mu 5}^{i-1},\ee where $a$ is the lattice size. In this way
the discretized version of our original 5-dimensional gauge theory
is substituted by an infinite collection of four-dimensional gauge
theories with gauge interacting chiral fields $\Sigma_i$ \be S=\int
d^4x\frac{a}{g_5^2}\left(-\frac 1 2\sum_i{\rm Tr}\left[F_{\mu\nu}^i
F^{\mu\nu i}\right]+\frac 1{a^2} {\rm
Tr}\left[\left(D_\mu\Sigma_i\right)\left(D^\mu\Sigma_i\right)^\dagger\right]\right).\ee
The theory obtained in this way is just an example of a larger set
of theories generically called "deconstructed theories"
\cite{Arkani-Hamed:2001ca} synthetically described by a moose
diagram (see Fig. \ref{fig:2}).

\section{2. Breaking the EW Symmetry without Higgs Fields}

\begin{figure}
  \includegraphics[height=.1\textheight]{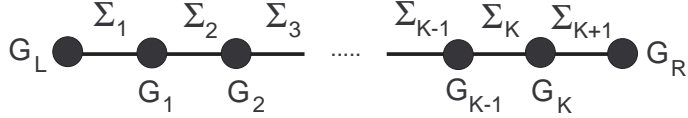}
  \caption{The simplest moose diagram for the Higgsless breaking of the EW symmetry.
  \label{fig:3}}
\end{figure}

As we have seen in the previous Section, abstracting from the
5-dimensional example one can study more general moose geometries.
The general structure will consist in many copies of the gauge group
$G$ intertwined by link variables $\Sigma$. Now suppose that we want
to describe the electro-weak (EW) symmetry breaking in this context.
The condition we have to satisfy is that, before the EW gauge group
is introduced, {\bf 3 massless Goldstone bosons should be present
(to give masses to $W^\pm$ and $Z$) and all the moose gauge fields
should be massive}. In the simplest case we take all the moose gauge
groups equal to $SU(2)$. Then, each $\Sigma$ field is an $SU(2)$
matrix \be
\Sigma_i=e^{i\vec\pi\cdot\vec\tau/(2f_i)},\label{eq:15}\ee with
$\vec \tau$ the Pauli matrices. Therefore each $\Sigma_i$ describes
three spin zero fields ($\pi_i$). In a connected moose diagram any
site (containing three gauge fields) may absorb one link (the 3
Goldstones $\pi_i$) giving rise to three massive gauge bosons.
Therefore our condition translates into \be number~of~
links=number~of~sites + 1.\ee The simplest of these  moose is the
"linear moose" whose diagram is given in Fig. \ref{fig:3}. The
corresponding action is \be S_{moose}=\int
d^4x\left(-\sum_{i=1}^K\frac 1{2}{\rm Tr}\left[F_{\mu\nu}^i
F^{\mu\nu i}\right]+\sum_{i=1}^{K+1}f_i^2{\rm
Tr}\left[\left(D_\mu\Sigma_i\right)\left(D^\mu\Sigma_i\right)^\dagger\right]\right).
\label{eq:17}\ee

We have now $K$ gauge groups $SU(2)$ and $K+1$ chiral fields. Notice
that the model has two global symmetries $G_L$ and $G_R$ associated
to the chiral fields $\Sigma_1$ and $\Sigma_{K+1}$ \be \Sigma_1\to
U_L\Sigma_1,~~~\Sigma_{K+1}\to \Sigma_{K+1}U_R^\dagger.\ee As such
they have been associated to the ends of the moose in Fig.
\ref{fig:3}. It is this global symmetry, $G_L\otimes G_R=
SU(2)_L\otimes SU(2)_R$, that is gauged by the standard group
$SU(2)_L\otimes U(1)$, in order to give the standard massive gauge
bosons $W^\pm$ and $Z$ and the massless photon. In fact, the three
Goldstones remaining after that the moose gauge fields have eaten up
the chiral fields are just the ones necessary for the breaking of
the EW symmetry. Prototypes of this theory are the BESS model for
$K=1$ \cite{Casalbuoni:1985kq} and its generalizations
\cite{Casalbuoni:1989xm}.

\section{3. EW Corrections for the Linear Moose}

If the moose vector fields are heavy enough it is possible to derive
an effective action describing only the Standard Model (SM) fields.
By denoting the typical mass of the moose vector fields by $M_V$, at
the leading order in $(M_W/M_V)^2$ one gets the usual SM relations
\be
M_W^2=\frac{v^2}4g^2,~~~M_Z^2=\frac{M_W^2}{c_\theta^2},~~~e=gs_\theta=g'c_\theta,\ee
with ($v\approx 250 ~GeV$)\be \frac 4{v^2}\equiv \frac
1{f^{\,2}}=\sum_{i=1}^{K+1}\frac 1{f_i^{,2}}.\label{eq:20}\ee In
this class of models all the corrections from new physics arise from
mixing of the SM vector bosons with the moose vector fields and
therefore are oblique corrections. As well known the oblique
corrections are completely captured by the parameters $S$, $T$ and
$U$ \cite{Peskin:1990zt,Peskin:1992sw} or, equivalently by the
parameters $\epsilon_i$, $i=1,2,3$
\cite{Altarelli:1991zd,Altarelli:1998et}. For the linear moose, the
existence of the global symmetry (custodial) $SU(2)_V$  ensures that
\be \epsilon_1=\epsilon_2=0,\ee or, equivalently $U=T=0$.

To compute the new physics contribution to the electroweak parameter
$\epsilon_3$ \cite{Altarelli:1991zd} we will make use of the
dispersive representation given in Refs.
\cite{Peskin:1990zt,Peskin:1992sw} for the related parameter $S$
($\epsilon_3=g^2 S/(16\pi)$)  \be
\epsilon_3=-\frac{g^2}{4\pi}\int_0^\infty\frac{ds}{s^2}
Im\left[\Pi_{VV}(s)-\Pi_{AA}(s)\right],\ee where $g$ is the
$SU(2)_L$ gauge coupling and $\Pi_{VV}(AA)$ is the current-current
correlator \be \int d^4x e^{-iq\cdot x}\langle
J^\mu_{V(A)}J^\nu_{V(A)}\rangle=ig^{\mu\nu}\Pi_{VV(AA)}(q^2)+(q^\mu
q^\nu~{\rm terms}).\ee $J_{V/A\mu}$ are the vector and axial
currents associated to the global symmetry $SU(2)_L\otimes SU(2)_R$,
getting the following contributions from the moose vector fields\bea
J_{V\mu}^a\Big|_{\rm vector~mesons}&=&f_1^2g_1A_{\mu}^{1a}+f_{K+1}^2g_KA_\mu^{Ka},\nn\\
J_{A\mu}^a\Big|_{\rm vector~
mesons}&=&f_1^2g_1A_{\mu}^{1a}-f_{K+1}^2g_KA_\mu^{Ka}.\eea It should
be noticed that the $\epsilon_3$ parameter is evaluated with
reference to the SM, and therefore the corresponding contributions
should be subtracted. For instance the contribution of the pion pole
to $\Pi_{AA}$, that is of the Goldstone particles giving mass to the
$W$ and $Z$ gauge bosons, does not appear in $\epsilon_3$. As
described previously, in the model described by the action
(\ref{eq:17}) all the new physics contribution comes from the new
vector bosons (we are assuming the standard couplings for the
fermions to $SU(2)_L\otimes U(1)$). Therefore from\be
Im\,\Pi_{VV(AA)}=-\pi\sum_{Vn,An}g^2_{nV,nA}\delta(s-m^2_n),\ee we
get \be
\epsilon_3=\frac{g^2}4\sum_n\left(\frac{g_{nV}^2}{m^4_n}-\frac{g^2_{nA}}{m^4_n}\right),\ee
where $g_{nV/A}$ are the decay coupling constants of the moose
vector fields defined by \be \langle 0|J_{V\mu}^a|\tilde
A^n_b(p,\epsilon)\rangle=g_{nV}\delta^{ab}\epsilon_\mu,~~~ \langle
0|J_{A\mu}^a|\tilde
A^n_b(p,\epsilon)\rangle=g_{nA}\delta^{ab}\epsilon_\mu,\ee and
$\tilde A^n_b(p,\epsilon)$ are the mass eigenstates of the moose
vector bosons. As shown in \cite{Casalbuoni:2004id} we can express
$\epsilon_3$ in two equivalent ways (see also
\cite{Hirn:2004ze,Georgi:2004iy})
 \be \epsilon_3=g^2g_1 g_K f_1^{\,2}
f_{K+1}^{\,2}(M_2^{-2})_{1K}=g^2\sum_{i=1}^K\frac{(1-y_i)y_i}{g_i^2},\label{eps3}
\ee where $M_2$ is the matrix of the square masses of the moose
vector bosons, and  \be y_i=\sum_{j=1}^i
x_j,~~~x_i=\frac{f^2}{f_i^2},~~~ \frac 1 {f^2}=\sum_{i=1}^{K+1}\frac
1 {f_i^2}~\Rightarrow~ \sum_{i=1}^{K+1}x_i=1. \label{f2}\ee Since
$0\le y_i\le 1$ it follows $\epsilon_3\ge 0$ (see also
\cite{Hirn:2004ze,Barbieri:2003pr,Chivukula:2004kg}). As an example,
let us take all the link couplings $f_i$ equal to a common value
$f_c$, and the same for the gauge couplings $g_i=g_c$. Then (see
also \cite{Foadi:2003xa})\be \epsilon_3=\frac 1
6\frac{g^2}{g_c^2}\frac{K(K+2)}{(K+1)}.\ee If we want to be
compatible with the experimental data we need to get
$\epsilon_3\approx 10^{-3}$. For $K=1$ this would require $g_c\ge
15.8 g$, implying a strong interacting gauge theory in the moose
sector. Notice also that insisting on a weak gauge theory would
imply $g_c$ of the order of $g$, let us say $g_c\approx 2\div 5 g$.
Then the natural value of $\epsilon_3$ would be of the order
$10^{-1}-10^{-2}$, incompatible with the experimental data.
\begin{figure}
  \includegraphics[height=.14\textheight]{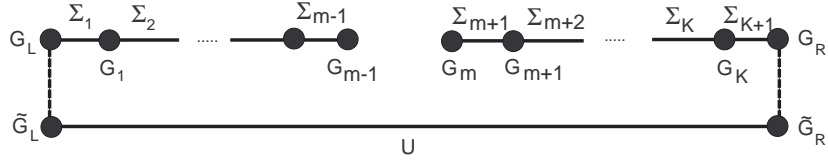}
  \caption{The diagram illustrates how a cut link model is generated from a linear moose.
  \label{fig:4}}
\end{figure}

Possible ways of evading the $\epsilon_3$ problem have been
considered in \cite{Casalbuoni:2004id}. A way is to cut a link, that
is to assume one of the link couplings, say $f_m$, equal to zero. In
this case the matrix of the mass square of the moose vector bosons
becomes block diagonal and, as a consequence, the same happens for
$M_2^{-2}$. Therefore $(M_2^{-2})_{1K}=0$ and $\epsilon_3=0$. Since,
suppressing a link amounts to eliminate three scalar fields, we need
a way to reintroduce them.  The $\Sigma_m$ field can be reintroduced
through a discretized version of a Wilson line \be
U=\Sigma_1\Sigma_2\cdots\Sigma_{K}\Sigma_{K+1},\ee and inserting in
the lagrangian a term \be f_0^2 Tr[
\partial_\mu U^\dagger \partial^\mu U].
\ee This term has a global invariance $\tilde G_L\otimes \tilde
G_R=SU(2)_L\otimes SU(2)_R$ originating from a transformation $U\to
\tilde U_L U\tilde U_R$. This invariance is different from the
original $G_L\otimes G_R$ before the EW gauging. As a consequence
the model has an enhanced custodial symmetry $\left[SU(2)_L\otimes
SU(2)_R\right]$ which is enough to ensure $\epsilon_3=0$
\cite{Inami:1992rb}. A particular example of this model, for $K=2$
($D-BESS$), was studied in
\cite{Casalbuoni:1995yb,Casalbuoni:1996qt} (originally introduced in
\cite{ Casalbuoni:1989xm}).

Another possibility \cite{Casalbuoni:2004id} is to suppress a link,
that is to assume a hierarchy among the links. As an example assume
an exponential behavior \be f_i=\bar f e^{c(i-1)},~~~~ g_i=g_c.\ee
\begin{figure}[h]
  \includegraphics[height=.22\textheight]{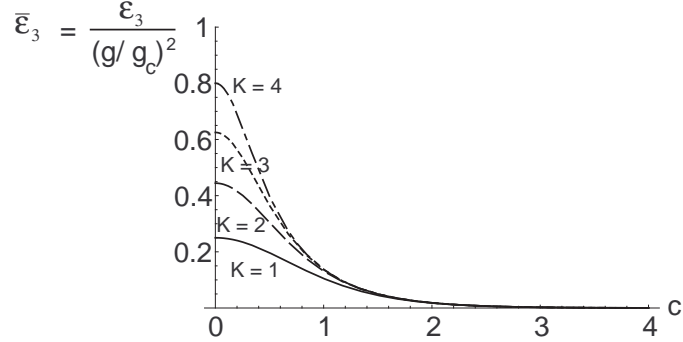}
  \caption{The behavior of $\epsilon_3$ (normalized to $(g/g_c)^2$) as a function of $c$
  in the exponentially suppressed linear model for different values of $K$.
  \label{fig:6}}
\end{figure}
From Fig. \ref{fig:6} we see that there is a big suppression factor,
of order $10^{-2}$ already for $c=2$. In fact expanding at the
leading order for large $c$ it is easily seen that \be
\epsilon_3\rightarrow \frac{g^2}{g_c^2}e^{-2c}.\ee However, lowering
or cutting the links may give rise to unitarity problems. For
instance, in the cut model $f_0$ must be of the order of the v.e.v.
of the Higgs field in the SM making the unitarity limit of these
class of model the same as in the SM without Higgs. We will study
the unitarity limits of the moose models in the next Section.

\section{4. Unitarity bounds for the linear moose}

The worst high-energy behavior of the moose models arises from the
scattering of longitudinal vector bosons. To simplify the
calculation we will make use of the equivalence theorem, that is of
the possibility of evaluating this amplitude in terms of the
scattering amplitude of the corresponding Goldstone bosons
\cite{Cornwall:1974rn}. However this theorem holds in the
approximation where the energy of the process is much higher of the
mass of the vector bosons. We will consider two situations. In the
first one we assume that all the moose vectors  have a mass,
$M_{V_i}$, much higher than the SM vector boson masses, in such a
way that we can evaluate the amplitude for the SM $W$ and $Z$ at
energies $M_{W/Z}\ll E\ll M_{V_i}$. The only Goldstone bosons of
interest here are the ones giving mass to $W$ and $Z$. The unitary
gauge for these bosons is given by the choice \be \Sigma_i=e^{i
f\vec\pi\cdot\vec\tau/(2f_i^2)},\ee with $f$ given in eq.
(\ref{eq:20}). The resulting four-pion amplitude is \be
A_{\pi^+\pi^-\to\pi^+\pi^-}=-\frac{f^4 u}4\sum_{i=1}^{K+1}\frac 1
{f_i^6}+\frac {f^4}4\sum_{i,j=1}^K
L_{ij}\left((u-t)(s-M_2)_{ij}^{-1}+(u-s)(t-M_2)_{ij}^{-1}\right),\ee
with \be L_{ij}=g_ig_j\left(\frac 1 {f_i^2}+\frac 1
{f_{i+1}^2}\right)\left(\frac 1 {f_j^2}+\frac 1
{f_{j+1}^2}\right).\ee This expression reproduces correctly the
low-energy limit, $E\ll M_{V_i}$: \be
A_{\pi^+\pi^-\to\pi^+\pi^-}\to-\frac{f^4
u}4\left(\sum_{i=1}^{K+1}\frac 1 {f_i^2}\right)^3=-\frac
{u}{4f^2}=-\frac u {v^2},\ee whereas in the high-energy limit, where
we can neglect the second term, \be
A_{\pi^+\pi^-\to\pi^+\pi^-}=-\frac{f^4 u}4\sum_{i=1}^{K+1}\frac 1
{f_i^6}.\ee The best unitarity limit is obtained for all the $f_i$'s
being equal to a common value $f_c$. In this case \be
A_{\pi^+\pi^-\to\pi^+\pi^-}=-\frac u{(K+1)v^2},\ee leading to the
unitarity bound \be \Lambda_{moose}=(K+1)\Lambda_{HSM}\approx 1.2
(K+1) ~TeV,\ee where $\Lambda_{HSM}$ is the unitary bound for the
Higgsless SM. In this case it is possible to improve as much as we
like the unitarity bound of the SM increasing $K$. However this
would lead to contradictions with the experimental bounds on
$\epsilon_3$.

As a second instance we consider an energy much higher than all the
masses of the vector bosons. In this case to determine the unitarity
bounds one has to consider the eigenchannel amplitudes corresponding
to all the possible four-longitudinal vector bosons. But, since the
unitary gauge for all the vector bosons is simply given by the
expression (\ref{eq:15}), the amplitudes are already diagonal, and
the result at high energy is simply \be
A_{\pi^+\pi^-\to\pi^+\pi^-}\to -\frac{u}{4f_i^2}.\ee We see that the
unitarity limit is determined the smallest link coupling. Therefore
in the exponentially suppressed model the unitarity bound is
essentially the same as in the SM, since in order to respect the
constraint given by the first equality in eq. (\ref{eq:20}), the
lowest coupling must be of order $v$. Also in this case the best
unitarity limit is for all the link couplings being equal $f_i=f_c$.
Then (for similar results see
\cite{Chivukula:2002ej})\be\Lambda_{moose}=\sqrt{K+1}\Lambda_{HSM}\approx
1.2 \sqrt{K+1}~TeV.\ee However, in order our approximation is
correct we have to require $M_{V_i}^{max}\ll \Lambda_{moose}$, and
since we expect roughly (assuming $g_c\approx g$)
$M_{V_i}^{max}\approx KM_W$, we get a bound $\sqrt{K}\ll 14$. By
taking $\sqrt{K}$ of order $2\div 3$ one could improve of the same
factor the SM unitarity bound, but again this would be hardly
compatible with the electro-weak experimental data.

\section{5. Delocalizing fermions}

\begin{figure}[h]
  \includegraphics[height=.27\textheight]{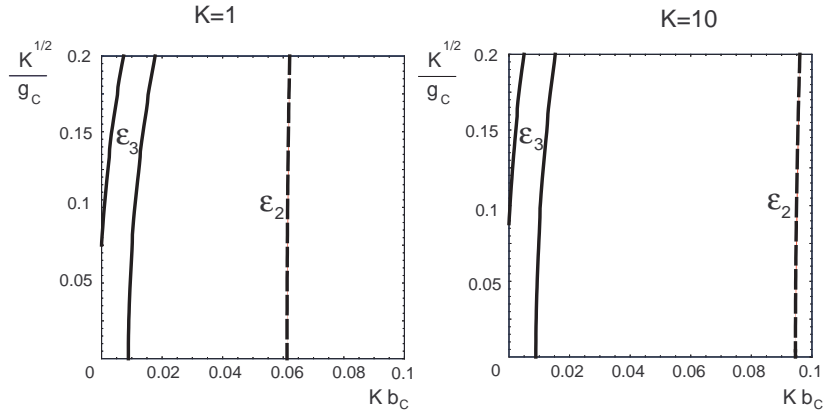}
  \caption{The 95\% C.L. allowed region in the plane $(Kb_c,\sqrt{K}/g_c)$ is
  the region on the left delimited by the two continuous lines coming form the
  bounds on $\epsilon_3$. The dashed line
  comes from a  bound on $\epsilon_2$, whereas the other
  bound form $\epsilon_2$ and the bounds from $\epsilon_1$ are out
  of the figure. The radiative
  corrections have been assumed as in the SM with $m_H= 1~TeV$ and
  $m_{top}=178~GeV$.
  \label{fig:7}}
\end{figure}
\begin{figure}[h]
  \includegraphics[height=.24\textheight]{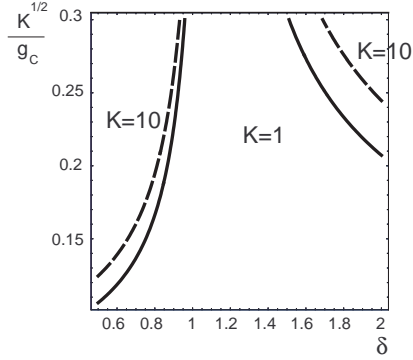}
  \caption{The 95\% C.L. allowed regions in the plane $(\delta,\sqrt{K}/g_c)$
  are the ones at the interior of the lines corresponding to $K=1$
  (continuous) and $K=10$ (dashed). The radiative corrections have
  been chosen as in Fig. \ref{fig:7}.
  \label{fig:8}}
\end{figure}

As we have seen, it is not possible to satisfy at the same time  the
experimental bounds on $\epsilon_3$ and improve in a sensible way
the unitarity limit. A way out has been considered in
\cite{Casalbuoni:2005rs,Chivukula:2005bn,SekharChivukula:2005xm}
allowing delocalized couplings of the SM fermions to the moose gauge
fields and some amount of fine tuning. In fact, the SM fermions can
be coupled to any of the gauge fields staying at the lattice sites
by means of a Wilson line.  However, we will consider only
left-handed fermions, since analogous interactions for the
right-handed ones are very much constrained
\cite{Casalbuoni:1985kq,Casalbuoni:1986vq}. Define \be
\chi_L^i=\Sigma_i^\dagger\Sigma_{i-1}^\dagger\cdots\Sigma_1^\dagger\psi_L.\ee
Then, under a gauge transformation, $\chi_L^i\to U_i\chi_L^i$, with
$U_i\in G_i$. We see that at each site we can introduce a gauge
invariant coupling given by \be
b_i\bar\chi_L^i\gamma^\mu\left(\de_\mu+ig_i A_\mu^i+\frac i 2
g'(B-L)Y_\mu\right)\chi_L^i.\ee The expressions for the parameters
$\epsilon_i$ are modified, and at first order in the couplings $b_i$
we get \be \epsilon_1\approx 0,~~~\epsilon_2\approx
0,~~~\epsilon_3\approx\sum_{i=1}^K
y_i\left(\frac{g^2}{g_i^2}(1-y_i)-b_i\right).\ee Therefore, with
some amount of fine tuning is possible to agree with the
electro-weak experimental data. To show it, let us take again all
the link couplings equal to $f_c$ and the gauge couplings equal to
$g_c$. We have considered two possibilities. In the first one we
take also the $b_i$ equal to a common value $b_c$. Then the allowed
region in the space $(Kb_c,\sqrt{K}/g_c)$ (we have chosen these
parameters due to the scaling properties of $g_c$ and $b$ with $K$)
is given in Fig. \ref{fig:7}.

In the second case we require a sort of local cancelation, assuming
(again $g_i=g_c$, $f_i=f_c$) \be b_i=\delta\frac{g^2}{g_i^2}(1-y_i)=
\delta\frac{g^2}{g_c^2}\left(1-\frac i {K+1}\right).\ee The allowed
region in the space $(\delta,\sqrt{K}/g_c)$ is given in Fig.
\ref{fig:8}. In this way it is possible to satisfy the EW
constraints and improve the unitarity bound of the Higgsless SM at
the same time.


\begin{thebibliography}{27}
\expandafter\ifx\csname
natexlab\endcsname\relax\def\natexlab#1{#1}\fi
\expandafter\ifx\csname bibnamefont\endcsname\relax
  \def\bibnamefont#1{#1}\fi
\expandafter\ifx\csname bibfnamefont\endcsname\relax
  \def\bibfnamefont#1{#1}\fi
\expandafter\ifx\csname citenamefont\endcsname\relax
  \def\citenamefont#1{#1}\fi
\expandafter\ifx\csname url\endcsname\relax
  \def\url#1{\texttt{#1}}\fi
\expandafter\ifx\csname urlprefix\endcsname\relax\def\urlprefix{URL
}\fi \providecommand{\bibinfo}[2]{#2}
\providecommand{\eprint}[2][]{\url{#2}}

\bibitem[{\citenamefont{Arkani-Hamed et~al.}(1998)\citenamefont{Arkani-Hamed,
  Dimopoulos, and Dvali}}]{Arkani-Hamed:1998rs}
\bibinfo{author}{\bibfnamefont{N.}~\bibnamefont{Arkani-Hamed}},
  \bibinfo{author}{\bibfnamefont{S.}~\bibnamefont{Dimopoulos}},
  \bibnamefont{and} \bibinfo{author}{\bibfnamefont{G.~R.} \bibnamefont{Dvali}},
  \bibinfo{journal}{Phys. Lett.} \textbf{\bibinfo{volume}{B429}},
  \bibinfo{pages}{263} (\bibinfo{year}{1998}), \eprint{hep-ph/9803315}.

\bibitem[{\citenamefont{Antoniadis et~al.}(1998)\citenamefont{Antoniadis,
  Arkani-Hamed, Dimopoulos, and Dvali}}]{Antoniadis:1998ig}
\bibinfo{author}{\bibfnamefont{I.}~\bibnamefont{Antoniadis}},
  \bibinfo{author}{\bibfnamefont{N.}~\bibnamefont{Arkani-Hamed}},
  \bibinfo{author}{\bibfnamefont{S.}~\bibnamefont{Dimopoulos}},
  \bibnamefont{and} \bibinfo{author}{\bibfnamefont{G.~R.} \bibnamefont{Dvali}},
  \bibinfo{journal}{Phys. Lett.} \textbf{\bibinfo{volume}{B436}},
  \bibinfo{pages}{257} (\bibinfo{year}{1998}), \eprint{hep-ph/9804398}.

\bibitem[{\citenamefont{Hill et~al.}(2001)\citenamefont{Hill, Pokorski, and
  Wang}}]{Hill:2000mu}
\bibinfo{author}{\bibfnamefont{C.~T.} \bibnamefont{Hill}},
  \bibinfo{author}{\bibfnamefont{S.}~\bibnamefont{Pokorski}}, \bibnamefont{and}
  \bibinfo{author}{\bibfnamefont{J.}~\bibnamefont{Wang}},
  \bibinfo{journal}{Phys. Rev.} \textbf{\bibinfo{volume}{D64}},
  \bibinfo{pages}{105005} (\bibinfo{year}{2001}), \eprint{hep-th/0104035}.

\bibitem[{\citenamefont{Cheng et~al.}(2001)\citenamefont{Cheng, Hill, Pokorski,
  and Wang}}]{Cheng:2001vd}
\bibinfo{author}{\bibfnamefont{H.-C.} \bibnamefont{Cheng}},
  \bibinfo{author}{\bibfnamefont{C.~T.} \bibnamefont{Hill}},
  \bibinfo{author}{\bibfnamefont{S.}~\bibnamefont{Pokorski}}, \bibnamefont{and}
  \bibinfo{author}{\bibfnamefont{J.}~\bibnamefont{Wang}},
  \bibinfo{journal}{Phys. Rev.} \textbf{\bibinfo{volume}{D64}},
  \bibinfo{pages}{065007} (\bibinfo{year}{2001}), \eprint{hep-th/0104179}.

\bibitem[{\citenamefont{Arkani-Hamed
  et~al.}(2001{\natexlab{a}})\citenamefont{Arkani-Hamed, Cohen, and
  Georgi}}]{Arkani-Hamed:2001ca}
\bibinfo{author}{\bibfnamefont{N.}~\bibnamefont{Arkani-Hamed}},
  \bibinfo{author}{\bibfnamefont{A.~G.} \bibnamefont{Cohen}}, \bibnamefont{and}
  \bibinfo{author}{\bibfnamefont{H.}~\bibnamefont{Georgi}},
  \bibinfo{journal}{Phys. Rev. Lett.} \textbf{\bibinfo{volume}{86}},
  \bibinfo{pages}{4757} (\bibinfo{year}{2001}{\natexlab{a}}),
  \eprint{hep-th/0104005}.

\bibitem[{\citenamefont{Arkani-Hamed
  et~al.}(2001{\natexlab{b}})\citenamefont{Arkani-Hamed, Cohen, and
  Georgi}}]{Arkani-Hamed:2001nc}
\bibinfo{author}{\bibfnamefont{N.}~\bibnamefont{Arkani-Hamed}},
  \bibinfo{author}{\bibfnamefont{A.~G.} \bibnamefont{Cohen}}, \bibnamefont{and}
  \bibinfo{author}{\bibfnamefont{H.}~\bibnamefont{Georgi}},
  \bibinfo{journal}{Phys. Lett.} \textbf{\bibinfo{volume}{B513}},
  \bibinfo{pages}{232} (\bibinfo{year}{2001}{\natexlab{b}}),
  \eprint{hep-ph/0105239}.

\bibitem[{\citenamefont{Casalbuoni et~al.}(1985)\citenamefont{Casalbuoni,
  De~Curtis, Dominici, and Gatto}}]{Casalbuoni:1985kq}
\bibinfo{author}{\bibfnamefont{R.}~\bibnamefont{Casalbuoni}},
  \bibinfo{author}{\bibfnamefont{S.}~\bibnamefont{De~Curtis}},
  \bibinfo{author}{\bibfnamefont{D.}~\bibnamefont{Dominici}}, \bibnamefont{and}
  \bibinfo{author}{\bibfnamefont{R.}~\bibnamefont{Gatto}},
  \bibinfo{journal}{Phys. Lett.} \textbf{\bibinfo{volume}{B155}},
  \bibinfo{pages}{95} (\bibinfo{year}{1985}).

\bibitem[{\citenamefont{Casalbuoni et~al.}(1989)\citenamefont{Casalbuoni,
  De~Curtis, Dominici, Feruglio, and Gatto}}]{Casalbuoni:1989xm}
\bibinfo{author}{\bibfnamefont{R.}~\bibnamefont{Casalbuoni}},
  \bibinfo{author}{\bibfnamefont{S.}~\bibnamefont{De~Curtis}},
  \bibinfo{author}{\bibfnamefont{D.}~\bibnamefont{Dominici}},
  \bibinfo{author}{\bibfnamefont{F.}~\bibnamefont{Feruglio}}, \bibnamefont{and}
  \bibinfo{author}{\bibfnamefont{R.}~\bibnamefont{Gatto}},
  \bibinfo{journal}{Int. J. Mod. Phys.} \textbf{\bibinfo{volume}{A4}},
  \bibinfo{pages}{1065} (\bibinfo{year}{1989}).

\bibitem[{\citenamefont{Peskin and Takeuchi}(1990)}]{Peskin:1990zt}
\bibinfo{author}{\bibfnamefont{M.~E.} \bibnamefont{Peskin}} \bibnamefont{and}
  \bibinfo{author}{\bibfnamefont{T.}~\bibnamefont{Takeuchi}},
  \bibinfo{journal}{Phys. Rev. Lett.} \textbf{\bibinfo{volume}{65}},
  \bibinfo{pages}{964} (\bibinfo{year}{1990}).

\bibitem[{\citenamefont{Peskin and Takeuchi}(1992)}]{Peskin:1992sw}
\bibinfo{author}{\bibfnamefont{M.~E.} \bibnamefont{Peskin}} \bibnamefont{and}
  \bibinfo{author}{\bibfnamefont{T.}~\bibnamefont{Takeuchi}},
  \bibinfo{journal}{Phys. Rev.} \textbf{\bibinfo{volume}{D46}},
  \bibinfo{pages}{381} (\bibinfo{year}{1992}).

\bibitem[{\citenamefont{Altarelli and Barbieri}(1991)}]{Altarelli:1991zd}
\bibinfo{author}{\bibfnamefont{G.}~\bibnamefont{Altarelli}} \bibnamefont{and}
  \bibinfo{author}{\bibfnamefont{R.}~\bibnamefont{Barbieri}},
  \bibinfo{journal}{Phys. Lett.} \textbf{\bibinfo{volume}{B253}},
  \bibinfo{pages}{161} (\bibinfo{year}{1991}).

\bibitem[{\citenamefont{Altarelli et~al.}(1998)\citenamefont{Altarelli,
  Barbieri, and Caravaglios}}]{Altarelli:1998et}
\bibinfo{author}{\bibfnamefont{G.}~\bibnamefont{Altarelli}},
  \bibinfo{author}{\bibfnamefont{R.}~\bibnamefont{Barbieri}}, \bibnamefont{and}
  \bibinfo{author}{\bibfnamefont{F.}~\bibnamefont{Caravaglios}},
  \bibinfo{journal}{Int. J. Mod. Phys.} \textbf{\bibinfo{volume}{A13}},
  \bibinfo{pages}{1031} (\bibinfo{year}{1998}), \eprint{hep-ph/9712368}.

\bibitem[{\citenamefont{Casalbuoni et~al.}(2004)\citenamefont{Casalbuoni,
  De~Curtis, and Dominici}}]{Casalbuoni:2004id}
\bibinfo{author}{\bibfnamefont{R.}~\bibnamefont{Casalbuoni}},
  \bibinfo{author}{\bibfnamefont{S.}~\bibnamefont{De~Curtis}},
  \bibnamefont{and} \bibinfo{author}{\bibfnamefont{D.}~\bibnamefont{Dominici}},
  \bibinfo{journal}{Phys. Rev.} \textbf{\bibinfo{volume}{D70}},
  \bibinfo{pages}{055010} (\bibinfo{year}{2004}), \eprint{hep-ph/0405188}.

\bibitem[{\citenamefont{Hirn and Stern}(2004)}]{Hirn:2004ze}
\bibinfo{author}{\bibfnamefont{J.}~\bibnamefont{Hirn}} \bibnamefont{and}
  \bibinfo{author}{\bibfnamefont{J.}~\bibnamefont{Stern}},
  \bibinfo{journal}{Eur. Phys. J.} \textbf{\bibinfo{volume}{C34}},
  \bibinfo{pages}{447} (\bibinfo{year}{2004}), \eprint{hep-ph/0401032}.

\bibitem[{\citenamefont{Georgi}(2005)}]{Georgi:2004iy}
\bibinfo{author}{\bibfnamefont{H.}~\bibnamefont{Georgi}},
  \bibinfo{journal}{Phys. Rev.} \textbf{\bibinfo{volume}{D71}},
  \bibinfo{pages}{015016} (\bibinfo{year}{2005}), \eprint{hep-ph/0408067}.

\bibitem[{\citenamefont{Barbieri et~al.}(2004)\citenamefont{Barbieri, Pomarol,
  and Rattazzi}}]{Barbieri:2003pr}
\bibinfo{author}{\bibfnamefont{R.}~\bibnamefont{Barbieri}},
  \bibinfo{author}{\bibfnamefont{A.}~\bibnamefont{Pomarol}}, \bibnamefont{and}
  \bibinfo{author}{\bibfnamefont{R.}~\bibnamefont{Rattazzi}},
  \bibinfo{journal}{Phys. Lett.} \textbf{\bibinfo{volume}{B591}},
  \bibinfo{pages}{141} (\bibinfo{year}{2004}), \eprint{hep-ph/0310285}.

\bibitem[{\citenamefont{Chivukula et~al.}(2004)\citenamefont{Chivukula,
  Kurachi, and Tanabashi}}]{Chivukula:2004kg}
\bibinfo{author}{\bibfnamefont{R.~S.} \bibnamefont{Chivukula}},
  \bibinfo{author}{\bibfnamefont{M.}~\bibnamefont{Kurachi}}, \bibnamefont{and}
  \bibinfo{author}{\bibfnamefont{M.}~\bibnamefont{Tanabashi}},
  \bibinfo{journal}{JHEP} \textbf{\bibinfo{volume}{06}}, \bibinfo{pages}{004}
  (\bibinfo{year}{2004}), \eprint{hep-ph/0403112}.

\bibitem[{\citenamefont{Foadi et~al.}(2004)\citenamefont{Foadi, Gopalakrishna,
  and Schmidt}}]{Foadi:2003xa}
\bibinfo{author}{\bibfnamefont{R.}~\bibnamefont{Foadi}},
  \bibinfo{author}{\bibfnamefont{S.}~\bibnamefont{Gopalakrishna}},
  \bibnamefont{and} \bibinfo{author}{\bibfnamefont{C.}~\bibnamefont{Schmidt}},
  \bibinfo{journal}{JHEP} \textbf{\bibinfo{volume}{03}}, \bibinfo{pages}{042}
  (\bibinfo{year}{2004}), \eprint{hep-ph/0312324}.

\bibitem[{\citenamefont{Inami et~al.}(1992)\citenamefont{Inami, Lim, and
  Yamada}}]{Inami:1992rb}
\bibinfo{author}{\bibfnamefont{T.}~\bibnamefont{Inami}},
  \bibinfo{author}{\bibfnamefont{C.~S.} \bibnamefont{Lim}}, \bibnamefont{and}
  \bibinfo{author}{\bibfnamefont{A.}~\bibnamefont{Yamada}},
  \bibinfo{journal}{Mod. Phys. Lett.} \textbf{\bibinfo{volume}{A7}},
  \bibinfo{pages}{2789} (\bibinfo{year}{1992}).

\bibitem[{\citenamefont{Casalbuoni et~al.}(1995)}]{Casalbuoni:1995yb}
\bibinfo{author}{\bibfnamefont{R.}~\bibnamefont{Casalbuoni}}
  \bibnamefont{et~al.}, \bibinfo{journal}{Phys. Lett.}
  \textbf{\bibinfo{volume}{B349}}, \bibinfo{pages}{533} (\bibinfo{year}{1995}),
  \eprint{hep-ph/9502247}.

\bibitem[{\citenamefont{Casalbuoni et~al.}(1996)}]{Casalbuoni:1996qt}
\bibinfo{author}{\bibfnamefont{R.}~\bibnamefont{Casalbuoni}}
  \bibnamefont{et~al.}, \bibinfo{journal}{Phys. Rev.}
  \textbf{\bibinfo{volume}{D53}}, \bibinfo{pages}{5201} (\bibinfo{year}{1996}),
  \eprint{hep-ph/9510431}.

\bibitem[{\citenamefont{Cornwall et~al.}(1974)\citenamefont{Cornwall, Levin,
  and Tiktopoulos}}]{Cornwall:1974rn}
\bibinfo{author}{\bibfnamefont{J.~M.} \bibnamefont{Cornwall}},
  \bibinfo{author}{\bibfnamefont{D.~M.} \bibnamefont{Levin}}, \bibnamefont{and}
  \bibinfo{author}{\bibfnamefont{G.}~\bibnamefont{Tiktopoulos}},
  \bibinfo{journal}{Phys. Rev.} \textbf{\bibinfo{volume}{D11}},
  \bibinfo{pages}{1145} (\bibinfo{year}{1974}).

\bibitem[{\citenamefont{Chivukula and He}(2002)}]{Chivukula:2002ej}
\bibinfo{author}{\bibfnamefont{R.~S.} \bibnamefont{Chivukula}}
  \bibnamefont{and} \bibinfo{author}{\bibfnamefont{H.-J.} \bibnamefont{He}},
  \bibinfo{journal}{Phys. Lett.} \textbf{\bibinfo{volume}{B532}},
  \bibinfo{pages}{121} (\bibinfo{year}{2002}), \eprint{hep-ph/0201164}.

\bibitem[{\citenamefont{Casalbuoni et~al.}(2005)\citenamefont{Casalbuoni,
  De~Curtis, Dolce, and Dominici}}]{Casalbuoni:2005rs}
\bibinfo{author}{\bibfnamefont{R.}~\bibnamefont{Casalbuoni}},
  \bibinfo{author}{\bibfnamefont{S.}~\bibnamefont{De~Curtis}},
  \bibinfo{author}{\bibfnamefont{D.}~\bibnamefont{Dolce}}, \bibnamefont{and}
  \bibinfo{author}{\bibfnamefont{D.}~\bibnamefont{Dominici}},
  \bibinfo{journal}{Phys. Rev.} \textbf{\bibinfo{volume}{D71}},
  \bibinfo{pages}{075015} (\bibinfo{year}{2005}), \eprint{hep-ph/0502209}.

\bibitem[{\citenamefont{Chivukula
  et~al.}(2005{\natexlab{a}})\citenamefont{Chivukula, Simmons, He, Kurachi, and
  Tanabashi}}]{Chivukula:2005bn}
\bibinfo{author}{\bibfnamefont{R.~S.} \bibnamefont{Chivukula}},
  \bibinfo{author}{\bibfnamefont{E.~H.} \bibnamefont{Simmons}},
  \bibinfo{author}{\bibfnamefont{H.-J.} \bibnamefont{He}},
  \bibinfo{author}{\bibfnamefont{M.}~\bibnamefont{Kurachi}}, \bibnamefont{and}
  \bibinfo{author}{\bibfnamefont{M.}~\bibnamefont{Tanabashi}},
  \bibinfo{journal}{Phys. Rev.} \textbf{\bibinfo{volume}{D71}},
  \bibinfo{pages}{115001} (\bibinfo{year}{2005}{\natexlab{a}}),
  \eprint{hep-ph/0502162}.

\bibitem[{\citenamefont{Chivukula
  et~al.}(2005{\natexlab{b}})\citenamefont{Chivukula, Simmons, He, Kurachi, and
  Tanabashi}}]{SekharChivukula:2005xm}
\bibinfo{author}{\bibfnamefont{R.~S.} \bibnamefont{Chivukula}},
  \bibinfo{author}{\bibfnamefont{E.~H.} \bibnamefont{Simmons}},
  \bibinfo{author}{\bibfnamefont{H.-J.} \bibnamefont{He}},
  \bibinfo{author}{\bibfnamefont{M.}~\bibnamefont{Kurachi}}, \bibnamefont{and}
  \bibinfo{author}{\bibfnamefont{M.}~\bibnamefont{Tanabashi}},
  \bibinfo{journal}{Phys. Rev.} \textbf{\bibinfo{volume}{D72}},
  \bibinfo{pages}{015008} (\bibinfo{year}{2005}{\natexlab{b}}),
  \eprint{hep-ph/0504114}.

\bibitem[{\citenamefont{Casalbuoni et~al.}(1987)\citenamefont{Casalbuoni,
  De~Curtis, Dominici, and Gatto}}]{Casalbuoni:1986vq}
\bibinfo{author}{\bibfnamefont{R.}~\bibnamefont{Casalbuoni}},
  \bibinfo{author}{\bibfnamefont{S.}~\bibnamefont{De~Curtis}},
  \bibinfo{author}{\bibfnamefont{D.}~\bibnamefont{Dominici}}, \bibnamefont{and}
  \bibinfo{author}{\bibfnamefont{R.}~\bibnamefont{Gatto}},
  \bibinfo{journal}{Nucl. Phys.} \textbf{\bibinfo{volume}{B282}},
  \bibinfo{pages}{235} (\bibinfo{year}{1987}).

\end{thebibliography}


\end{document}